\overfullrule=0pt
\input harvmac

\def\a{\alpha}
\def\b{\beta}
\def\g{\gamma}
\def\l{\lambda}
\def\lh{\widehat\lambda}
\def\d{\delta}
\def\t{\theta}
\def\r{\rho}
\def\rh{\widehat\rho}
\def\s{\sigma}
\def\o{\omega}
\def\oh{\widehat\omega}
\def\O{\Omega}
\def\S{\Sigma}

\def\L{\Lambda}
\def\D{\Delta}
\def\N{\nabla}
\def\Nb{\overline\nabla}
\def\Pb{\overline\Pi}

\def\Sh{\widehat S}
\def\Ih{\widehat I}
\def\p{\partial}
\def\pb{\overline\partial}
\def\dh{\widehat d}
\def\Jb{\overline J}
\def\half{{1\over 2}}

\Title{ \vbox{\baselineskip12pt
\hbox{DFPD 04/TH/04}
\hbox{IFT-P.004/2004}}}
{\vbox{\centerline{ Conformal Invariance of the Pure Spinor Superstring}
\smallskip
\centerline{ in a Curved Background}}}

\smallskip
\centerline{Osvaldo Chand\'{\i}a\foot{e-mail:
chandia@pd.infn.it}}
\smallskip
\centerline{\it Dipartamento di Fisica, Universit\`a degli Studi di
Padova }
\centerline{\it
Istituto Nazionale di Fisica Nucleare, Sezione di Padova, Italia}
\bigskip
\centerline{Brenno Carlini Vallilo\foot{e-mail:
vallilo@ift.unesp.br}}
\smallskip
\centerline{\it
Instituto de F\'{\i}sica Te\'orica, Universidade
Estadual
Paulista}
\centerline{\it
Rua Pamplona 145, 01405-900, S\~ao Paulo, SP, Brasil}

\bigskip

\noindent
It is shown that the pure spinor formulation of the heterotic superstring
in a generic gravitational and super Yang-Mills background has vanishing
one-loop beta functions.

\Date{January 2004}

\newsec{Introduction}

It is a well known fact that the quantum mechanical preservation of the
conformal symmetry in string theory implies that the
background space-time satisfies suitable equations of motion. In the
simplest case, a
bosonic string coupled to a curved space-time geometry is described by a
conformal field theory if the space-time metric satisfies the Einstein
equations plus $\a'$-corrections.

The instabilities produced by the presence of the tachyon can be avoided
if
we include supersymmetry. There are two traditional ways to achieve this
end
in string theory. The first possibility considers susy at world-sheet
level,
so one constructs the so called Ramond-Neveu-Schwarz (RNS) superstring.
The second possibility adds
susy at the space-time level, so one describes the dynamics of the
superstring by using the Green-Schwarz (GS) model.

There are various problems in these approaches that lead to think in an
alternative description of the superstring. We mention some of them.
On the one hand, it is difficult to describe space-time fermions in the
RNS model and,
therefore, backgrounds with non trivial RR fields.
On the other hand, the lack of a covariant quantization of the GS
superstring
remains an open problem.

The alternative approach to describe a superstring is the recently
developed
Berkovits formalism \ref\berkovits{
N.~Berkovits,
``Super-Poincare Covariant Quantization of the Superstring,''
JHEP {\bf 0004}, 018 (2000)
[arXiv:hep-th/0001035].}. Here, the quantization is performed by
constructing
a BRST charge $Q=\int\l^\a d_\a$, where $\l^\a$ is a pure spinor\foot{It
is a
c-number field constrained to satisfy the pure spinor condition
$(\l\g^a\l)=0$ with $\g^a$ being the $16\times 16$ symmetric
ten-dimensional
gamma
matrices.} and $d_\a$ is the world-sheet generator of superspace
translations.
It has been possible to verify that the cohomology of $Q$ produces the
correct
superstring spectrum, in the light-cone gauge \ref\cohom{
N.~Berkovits,
``Cohomology in the Pure Spinor Formalism for the Superstring,''
JHEP {\bf 0009}, 046 (2000)
[arXiv:hep-th/0006003]\semi N.~Berkovits and O.~Chand\'{\i}a,
``Lorentz Invariance of the Pure Spinor BRST Cohomology for the
Superstring,''
Phys.\ Lett.\ B {\bf 514}, 394 (2001)
[arXiv:hep-th/0105149].}
and in a manifestly ten-dimensional super-Poincare covariant manner 
\ref\massive{
N.~Berkovits and O.~Chand\'{\i}a,
``Massive Superstring Vertex Operator in D = 10 Superspace,''
JHEP {\bf 0208}, 040 (2002)
[arXiv:hep-th/0204121].}. The model can be also formulated to describe the
coupling of the superstring in a background with RR fields, such as the
AdS$_5\times$S$^5$ \ref\ads{
N.~Berkovits and O.~Chand\'{\i}a,
``Superstring Vertex Operators in an AdS$_5\times$S$^5$ Background,''
Nucl.\ Phys.\ B {\bf 596}, 185 (2001)
[arXiv:hep-th/0009168].}, where one-loop conformal invariance has been
proved \ref\vallilo{
B.~C.~Vallilo, ``One-loop Conformal Invariance of the Superstring in an
AdS$_5\times$S$^5$ Background,''
JHEP {\bf 0212}, 042 (2002)
[arXiv:hep-th/0210064].}.

Let us be more precise in the case relevant for the present paper.
Consider
a pure spinor description of the heterotic superstring. It is shown in
\ref\bh{
N.~Berkovits and P.~S.~Howe,
 ``Ten-dimensional Supergravity Constraints from the Pure Spinor Formalism
for the Superstring,''
Nucl.\ Phys.\ B {\bf 635}, 75 (2002)
[arXiv:hep-th/0112160].} that the nilpotence of the BRST charge and the
holomorphicity of the BRST current put the background on-shell, at least
in
the classical level. The next step is to study this theory at the quantum
level. We fix the world-sheet surface to be a sphere, that is we are
considering string perturbation theory at the tree-level. But we can make
a quantum field theory for the sigma model and so construct a loop
expansion
on the sphere. In this way we can verify if the BRST charge remains
nilpotent
and if the BRST charge remains holomorphic. These conditions are related
to
the conformal invariance of the model at the quantum level. The argument
for
this is given in \bh\ for the linearized case. Of course, it would be
interesting to study the non-linear version of this. But we decide to
study
the
conformal invariance of the sigma model at the one-loop level by computing
the
beta functions. 

In an off-shell formulation of the N=1 supergravity \foot{See,
for example \ref\nilsson{
B.~E.~W.~Nilsson and A.~K.~Tollsten,
``The Geometrical Off-Shell Structure of Pure N=1 D = 10 Supergravity in
Superspace,''
Phys. Lett.\ B {\bf 169}, 369 (1986).}.} one might interpret the beta 
functions of 
the first two couplings in the sigma model action of (2.1) as the 
equations of motion for 
 two independent off-shell superfields which solve the supergravity
Bianchi indentities. On the SYM side one cannot find such coupling 
since there is no known off-shell superfield which solves the SYM Bianchi
identities. Since the constrains found by Berkovits and Howe already put 
the supergravity/SYM system on-shell all the superfields in the background 
field expansion done in this work are also on-shell.

In section 2 we will review the pure spinor sigma model for the heterotic
superstring in a generic background field and the constraints imposed
by the nilpotence of the BRST charge and the holomorphicity of
the BRST current. In section 3 we will perform a covariant
background field expansion of the sigma model action and in section 4 will
compute the one-loop effective action. In section 5 we will show that the
one-loop beta functions vanish due the classical superspace constraints of
\bh.

\newsec{The Pure Spinor Sigma Model}

Let us consider an heterotic string in a curved background. The action in
the pure spinor formalism is given by

\eqn\action{\eqalign{S={1\over{2\pi\a'}}&\int
d^2z~[\half\Pi^a\Pb^b\eta_{ab}+
\half\Pi^A\Pb^B B_{BA}+\Pi^A\Jb^I A_{IA}\cr
+d_\a(\Pb^\a+\Jb^I W^\a_I)&+\l^\a\o_\b(\Pb^A\O_{A\a}{}^\b+\Jb^I
U_{I\a}{}^\b)]+S_{\Jb}+S_{\l,\o}+S_{FT},\cr}}
where $\Pi^A=\p Z^M E_M{}^A, \Pb^A=\pb Z^M E_M{}^A$ with $E_M{}^A$ being
the supervielbein, $Z^M=(x^m,\t^\mu); m=0,\dots, 9, \mu=1,\dots, 16$ the
superspace coordinates, $d_\a$ the world-sheet generator of superspace
translations. $S_{\Jb}$ is the action for the gauge group variables,
$S_{\l,\o}$ is the action for the pure spinor variables $(\l,\o)$.
$S_{FT}$
is the Fradkin-Tseytlin term which is given by

\eqn\ft{S_{FT} = {1\over{2\pi}}\int d^2z~r~\Phi,}
where $r$ is the world-sheet scalar curvature and $\Phi$ is a superfield
whose $\t$-independent part is the dilaton. Recall that the
Fradkin-Tseytlin
term breaks conformal invariance, then it can be seen as an
$\a'$-correction
which restores it at the quantum level.

Since the
pure spinor variables can only enter in the combinations
$J=\l^\a\o_\a$ (the ghost number current) and
$N^{ab}=\half(\l\g^{ab}\o)$ (the generator for Lorentz rotations of the
pure spiror fields), the couplings between the
pure spinor variables and the background fields can be written as

\eqn\JN{\eqalign{\l^\a\o_\b\Jb^I U_{I\a}{}^\b&=J\Jb^I U_I
+ \half N^{ab}\Jb^IU_{Iab},\cr
\l^\a\o_\b\Pb^A \O_{A\a}{}^\b&=J\Pb^A \O_A+ \half
N^{ab}\Pb^A\O_{Aab}.\cr}}

The quantization of this system is determined by studying the BRST
operator $Q=\oint \l^\a d_\a$. In a flat background it is easy to show
that $Q^2=0$ and $\pb(\l^\a d_\a)=0$. In a curved background, the
nilpotence of the
BRST charge and holomorphicity of the BRST current at the classical level
 imply that the
background fields satisfy the N=1 SUGRA/SYM equations
of motion \bh. In fact, the nilpotence determines the constraints

\eqn\nilp{\l^\a\l^\b T_{\a\b}{}^A=\l^\a\l^\b H_{\a\b A}=\l^\a\l^\b\l^\g
R_{\a\b\g}{}^\d=\l^\a\l^\b F_{I\a\b}=0,}
and from the holomorphicity of the BRST current we obtain

\eqn\hol{\eqalign{T_{\a(ab)}=H_{\a ab}=T_{a\a}{}^\b&
=\l^\a\l^\b R_{a\a\b}{}^\g=0,\cr
T_{\a\b a}=H_{\a\b a},\quad F_{Ia\a}&=W^\b_I T_{\a\b a},\quad
F_{I\a\b}=\half
H_{\a\b\g}W^\g_I,\cr
\N_\a W^\b_I=U_{I\a}{}^\b-W^\g_I T_{\g\a}{}^\b,\quad &\l^\a\l^\b(\N_\a
U_{I\b}{}^\g+R_{\d\a\b}{}^\g W^\d_I)=0,\cr}}
where $T, R, H$ and $F$ are defined as follows. The covariant
derivative acting on a super p-form with values in the Lie algebra of
the gauge group is given by\foot{Wedge product between the superfields
is assumed.}

\eqn\deriv{\N\Psi^A=d\Psi^A+\Psi^B \O_B{}^A-\Psi^A A + (-1)^p A\Psi^A,}
where $d$ is the exterior derivative which maps a super p-form into a
super (p+1)-form. Note that by acting one more time the covariant
derivative
on $\Psi^A$ it is obtained

\eqn\nn{\N\N\Psi^A=\Psi^B R_B{}^A+F\Psi^A-\Psi^A F,}
where the curvatures $R$ and $F$ are given by

\eqn\curv{R_A{}^B=d\O_A{}^B+\O_A{}^C\O_C{}^B=\half E^D E^C R_{CDA}{}^B,}
$$
F=dA-AA=\half E^B E^A F_{IAB} K^I.$$
We can also define the three-form field strength
$$
H=dB={1\over 6}E^C E^B E^A H_{ABC},$$
where $E^A$ is the vielbein one form, the potential one form is $A=A_I
K^I$ with $[K^I, K^J]=f^{IJ}{}_L K^L$ ($f$'s are structure constants of
the Lie group). It is also necessary to define
the torsion 2-form

\eqn\torsion{T^A=\N E^A=\half E^C E^B T_{BC}{}^A.}

The curvatures and the torsion are constrained to satisfy the Bianchi
identities, $\N T^A=E^B R_B{}^A, dH=\N F=\N R_A{}^B=0$. In components
they read

\eqn\bianchi{\eqalign{(\N T)_{ABC}{}^D&\equiv\N_{[A}
T_{BC]}{}^D+T_{[AB}{}^E
T_{EC]}{}^D-R_{[ABC]}{}^D=0,\cr
(\N F)_{ABC}&\equiv\N_{[A} F_{BC]}+T_{[AB}{}^D F_{DC]}=0,\cr
(\N R)_{ABC}{}_{D}{}^E&\equiv\N_{[A}R_{BC]D}{}^E +
T_{[AB}{}^FR_{FC]D}{}^E=0,\cr
(\N H)_{ABCD}&\equiv\N_{[A} H_{BCD]}+{3\over 2} T_{[AB}{}^E
H_{ECD]}=0,\cr}}
where the antisymmetrization is over $A,B, C$ indices in the first three
identities, while in the fourth it is on $A, B, C, D$ indices.

It is also noted in \bh\ that the action \action\ has two independent
local Lorentz invariances. One acts on the bosonic indices of the vielbein
as $\d\Pi^a=\Pi^b \L_b{}^a$ and the other one acts on the fermionic
indices
of the vielbein as $\d\Pi^\a=\Pi^\b\ \S_\b{}^\a$ with
$\O_{A\a}{}^\b$ transforming as a connection under $\S$ and the
remaining variables of the action as covariant objects
(e.g. $\d\l^\a=\l^\b \S_\b{}^\a$). The action is also invariant under the
shift symmetry

$$
\d\O_\a=(\g_a)_{\a\b}h^{a\b},\quad
\d\O_\a{}^{ab}=2(\g^{[a})_{\a\b}h^{b]\b},
\quad \d d_\a=\d\O_{\a\b}{}^\g \l^\b\o_\g, \quad \d U_{I\a}{}^\b=W_I^\g
\d\O_{\g\a}{}^\b.
$$
With the help of these invariances, the constraints \nilp, \hol\ and
the Bianchi identities \bianchi\ it is possible to set
$H_{\a\b\g}=T_{\a\b}{}^\g=0$ and $T_{\a\b}{}^a=\g^a_{\a\b}$. It is argued
in
\bh\ that these constraints and the Bianchi identities \bianchi\ imply the
SUGRA/SYM equations. We will show that these choices allow to check the
vanishing of the one-loop beta functions.

In order to describe correctly the degrees of freedom of the SUGRA/SYM
system, the spinorial derivative of the dilaton  superfield must be
proportional to the connection $\O_\a$. In fact,
the one-loop preservation of $\pb (\l^\a d_\a)=0$ implies \bh\

\eqn\dpom{\N_\a \Phi = 4 \O_\a,}
which will be crucial to vanish the one-loop beta functions in the
subsequent discussion.

\newsec{The Covariant Background Field Method}

We expand the action \action\
in a covariant way. We need to expand the superspace
variables, the ghosts and the gauge group variables.

\subsec{Superspace expansion}

We follow the background field method based on normal
coordinates (see \ref\dbs{
J.~de Boer and K.~Skenderis,
``Covariant Computation of the Low Energy Effective Action of the
Heterotic
Superstring,''
Nucl.\ Phys.\ B {\bf 481}, 129 (1996)
[arXiv:hep-th/9608078].} and references therein).

We define local coordinates around a certain point in
superspace $Z^M$ by
specifying the value of its tangent along certain
geodesics. Normal
coordinates are those in which the geodesics looks like a
straight
line. If we denote by $Y^M$ the value of the tangent to the
geodesics at
$Z^M$, then the geodesics points to $Z^M+Y^M$. As it was
shown in \dbs\ any
tensor defined at $Z+Y$ can be identified with the tensor
at $Z$ by

$$
T'=e^{Y^A\N_A} T,$$
which can be iteratively obtained by applying the operator
$\D$ on $T$
defined as

\eqn\trans{
\D T=[Y^A\N_A,T].}
Here we are using coordinates in the orthonormal frame. Recall
$Y^A=Y^M E_M{}^A$ and $\N_A=E_A{}^M\N_M$, where $\N_M$ is the
covariant
derivative containing the Christoffel symbol while $\N_A$
is the covariant
derivative containing the spin connection. As it was argued
in \dbs, $\D T$
is the parallel transportation of the tensor from $Z+Y$ to
$Z$ through the
geodesics. Note also that the geodesic equation can be written as
$\D Y^A = 0$.

In order to expand the action in powers of $Y$ we need to
know how the
operator $\D$ acts on the different fields of \action. If we choose as a
tensor $T$ the covariant derivative of a zero form and if we use \deriv\
we can read off the action of $\D$ on the vielbein and the connections.
The
result is

\eqn\varn{\eqalign{&\D E_A{}^M=-(\N_A Y^B+Y^CT_{CA}{}^B)E_B{}^M,\cr
&
\D\O_{A\a}{}^\b=-(\N_A Y^B+Y^C T_{CA}{}^B)\O_{B\a}{}^\b+Y^B
R_{BA\a}{}^\b,\cr
&
\D A_{IA}=-(\N_A Y^B +Y^C T_{CA}{}^B)A_{IB}+Y^BF_{IBA}.\cr}}
We need to perform the expansion of  $\Pi^A=\p Z^M E_M{}^A$. Note that
$\p Z^M$ is
annihilated by
$\D$ since it does not change under parallel
transportation. By inverting
the first equation in \varn, one can see that

$$
\D E_M{}^A=E_M{}^B(\N_B Y^A+Y^C T_{CB}{}^A),$$
therefore
\eqn\varpi{\D\Pi^A=\N Y^A-Y^B\Pi^CT_{CB}{}^A=\N Y^A+\Pi^B Y^C T_{CB}{}^A,}
where we have defined $\N Y^A=\Pi^B\N_B Y^A$. By doing the
same kind of
calculation we obtain

$$\D(\N Y^A)=-Y^D Y^C \Pi^B R_{BCD}{}^A.$$
Analogously, for the barred world-sheet fields we obtain

$$
\D\Pb^A=\Nb Y^A-Y^B\Pb^C T_{CB}{}^A=\Nb Y^A+\Pb^B Y^C T_{CB}{}^A,$$
$$
\D(\Nb Y^A)=-Y^D Y^C \Pb^B R_{BCD}{}^A,$$
where $\Nb Y^A=\Pb^B \N_B Y^A$. Note that $B_{AB}, W^\a_I, U_I$ and
$U^{ab}_I$ are tensors, then they can
be expanded following the rule \trans.

We assume that $d_\a$ is a fundamental field, its expansion is

$$
d=d_0+\dh,$$
where $d_0$ is the background value and $\dh$ is the
quantum
fluctuation. Since this expansion is independent of the
superspace
expansion, it is satisfied that $\D\dh=0$.

\subsec{Ghost and gauge group expansion}

As in the case of the $d$ world-sheet field, the
pure-spinor ghosts can be
treated as fluctuations of some background value

$$
\l=\l_0+\lh,\quad \o=\o_0+\oh,$$
again, this expansion is independent of the superspace
expansion, then it is satisfied $\D\lh=\D\oh=0$. We will not enter in the
details of the propagator for these fields. Since only the covariant
combinations $J$ and $N^{ab}$ enter in the action,
we can make an expansion of the form

$$
J=J_0+J_1+J_2,\quad N^{ab}=N^{ab}_0+N^{ab}_1+N^{ab}_2,$$
where each subscript represents the order in the quantum
fluctuations.
We are interested in OPE's which depend on $(J_0,N^{ab}_0)$. This is
because
the OPE terms independent of the fields have short distance behavior
like ${1\over (z-w)^2}$, which does not contribute to divergences at one
loop.
And the terms with quantum fluctuations do not enter in the effective
action.
The only non vanishing OPE of the type above is

\eqn\qopeg{N^{ab}_1(z) N^{cd}_1(w)\to
{1\over(z-w)}[-\eta^{a[c}N_0^{d]b}(w)+\eta^{b[c}N_0^{d]a}(w)].}

In the same way, the gauge current can be expanded as

$$
\Jb^I=\Jb_0^I+\Jb_1^I+\Jb_2^I.$$
As before, we need only the OPE\foot{This OPE can be obtained
by realizing the gauge group contribution to the sigma-model by antichiral
Majorana spinors $\r^{\cal A}$ (${\cal A}=1,\dots, 32$) and noting
$\Jb^I=\half K^I_{\cal AB}\r^{\cal A}\r^{\cal B}$, where $K^I$ are the
gauge group generators. Then we expand these fermions as $\r^{\cal A} =
\r_0^{\cal A} + \rh^{\cal A}$.}

\eqn\qopej{\Jb^I_1({\bar z}) \Jb^J_1({\bar w})\to{1\over({\bar z}-{\bar
w})}f^{IJ}{}_K\Jb^K_0({\bar w}).}

In the following we will drop the the $0$ subindex, then we will denote
$d_{0\a}$ as $d_\a$, $J_0$ as $J$, $N^{ab}_0$ as $N^{ab}$ and
$\Jb^I_0$ as $\Jb^I$.

\newsec{The One-loop Effective Action}

Now we perform the expansion of the action \action\ in the way described
in the previous section with the variables
${\cal Q}= (Y, \rh, \lh, \oh)$ as the quantum fluctuations.
The effective action will be obtained by integrating
out ${\cal Q}$. For the one-loop contribution to this action, it is
necessary go up to
second order in the
quantum fluctuations. It is not so difficult to see that up to second
order
the expansion of the action has the form (we will set $\a'=1$)

$$
S=S_0 + \Sh + \Ih,$$
where $S_0$ is the background value of the action,
\eqn\spp{\Sh={1\over{2\pi}} \int d^2z~\half\eta_{ab}\N Y^a \Nb Y^b +
\dh_\a\Nb Y^\a +
L_g,}
determines the propagators for the
quantum fluctuations as \qopeg, \qopej\
and
\eqn\opes{Y^a(z, \bar z) Y^b(w, \bar w)\to -\eta^{ab}\log
|z-w|^2,}
$$
\dh_\a(z) Y^\b(w)\to {\d^\b_\a\over(z-w)},$$
note that these terms come from the expansion of the $\Pi^a\Pb_a$,
$d_\a \Pb^\a$ and $L_g$ terms of \action. And

\eqn\ihh{\eqalign{\Ih={1\over{2\pi}} &\int d^2z~ Y^A Y^B E_{AB}^{(1,1)}
+Y^A\N Y^B C_{AB}^{(0,1)}+Y^A\Nb Y^B
D_{AB}^{(1,0)}+\dh_\a Y^A G_A^{\a(0,1)}\cr
&+\Jb^I_1 Y^A
I_{IA}^{(1,0)}
+\Jb^I_2 K_I^{(1,0)}
+J_1 Y^A L_A^{(0,1)}
+J_2 M^{(0,1)}+N^{ab}_1 Y^A
O_{Aab}^{(0,1)}\cr
&+N^{ab}_2
P_{ab}^{(0,1)}
+J_1\Jb^I_1Q_I^{(0,0)}+N^{ab}_1\Jb^I_1
R_{Iab}^{(0,0)}+\dh_\a \Jb^I_1 S_I^{\a(0,0)}+{\cal O}({\cal Q}^3),\cr}}
where the superfields $C,\dots, S$ depend on the background superfields
and they can
be obtained by using the expansions defined in the previous section, but
we will not need all of them. The superscripts in these
superfields indicate their conformal weights.

As we said before, the effective action is determined by integrating out
the variables ${\cal Q}$, that is

$$
e^{-S_{eff}}= e^{-S_0}\int D{\cal Q}~e^{-\Sh}
[ 1 - \Ih + \half \Ih^2 + \cdots].$$
Remember that the lost of conformal invariance comes from the UV
divergences
of the Feynman diagrams \ref\afm{
L.~Alvarez-Gaum\'e, D.~Z.~Freedman and S.~Mukhi,
 ``The Background Field Method and the Ultraviolet Structure of the
Supersymmetric Nonlinear Sigma Model,''
Annals Phys.\  {\bf 134}, 85 (1981).}. There are two types of diagrams
which lead to UV divergences in this path integration.
A tadpole diagram, which is
formed in the single contractions in $\Ih$ of this expansion, and a `fish'
diagram,
which is formed by double contractions in the $\Ih^2$ term.

It should be noticed that the effective action should be
given by a
conformal weight $(1,1)$ density. Therefore, the only terms
will
contribute to it are those formed by single contraction of
$Y$'s in the
term with $E_{AB}^{(1,1)}$ in the action and from double
contractions in
$Y$'s between $Y^A\N Y^B C_{AB}^{(0,1)}$ with $Y^A\Nb Y^B
D_{AB}^{(1,0)}$
and $Y^A\Nb Y^B D_{AB}^{(1,0)}$ with $\dh_\a Y^A
G_A^{\a(0,1)}$ terms in
the action. Note that the OPE's between the gauge and ghost currents
\qopeg, \qopej\ provide the right conformal weights. Then, the double
contractions between $\Jb^I_1 Y^A
I_{IA}^{(1,0)}$ with $\dh_\a \Jb^I_1 S_I^{\a(0,0)}$ and between
$N^{ab}_1\Jb^I_1
R_{Iab}^{(0,0)}$ with itself will also contribute.

\subsec{Computation of the one-loop UV divergence}

The single contraction between $Y$'s in the
term with $E_{AB}^{(1,1)}$ in the action leads to the divergence

$$
-{1\over{2\pi}}\int d^2z ~\eta^{ab} E^{(1,1)}_{ab}\log\L,$$
where $\L$ is the momentum cut-off\foot{One can show that
$\log|0|^2=-\log\L$. It is also useful to know that
$\int d^2z/|z|^2=2\pi\log\L$.}. The double contraction between $Y^A\N
Y^B C_{AB}^{(0,1)}$ with $Y^A\Nb Y^B D_{AB}^{(1,0)}$ leads to

$$
{1\over{2\pi}}\int d^2z ~C_{ab}^{(0,1)} D_{cd}^{(1,0)}
\eta^{a[c}\eta^{d]b}\log\L.$$
The double contraction between $Y^A\Nb Y^B D_{AB}^{(1,0)}$ with $\dh_\a
Y^A
G_A^{\a(0,1)}$ is

$$
{1\over{2\pi}}\int d^2z~ \eta^{ab}(D_{\a a}^{(1,0)} -
D_{a \a}^{(1,0)}) G_b^{\a(0,1)}\log\L.$$
The double contraction between $\Jb^I_1 Y^A
I_{IA}^{(1,0)}$ with $\dh_\a \Jb^I_1 S_I^{\a(0,0)}$ gives

$$
-{1\over{2\pi}}\int d^2z~ \Jb^I f^{JK}{}_I I^{(1,0)}_{J\a}
S^{\a(0,0)}_K\log\L.$$
The double contraction between $N^{ab}_1\Jb^I_1
R_{Iab}^{(0,0)}$ with itself is

$$
-{1\over{2\pi}}\int d^2z~ \Jb^I N^{ab} f^{JK}{}_I
R_{Jc[a}^{(0,0)} R^{(0,0)}_{Kb]}{}^c\log\L.$$
In summary, the one loop divergence coming from integrating out the
quantum fluctuations is
\eqn\res{\eqalign{
S_\L=&{1\over{2\pi}}\int d^2z~[-\eta^{ab}
E^{(1,1)}_{ab}+\eta^{a[c}\eta^{d]b}
C_{ab}^{(0,1)} D_{cd}^{(1,0)} +\eta^{ab}D_{[\a
a]}^{(1,0)}G^{\a(0,1)}_b\cr
&-\Jb^I f^{JK}{}_I I^{(1,0)}_{J\a}
S^{\a(0,0)}_K-\Jb^I N^{ab} f^{JK}{}_I
R_{Jc[a}^{(0,0)} R^{(0,0)}_{Kb]}{}^c]\log\L.\cr}}

\subsec{The one-loop beta functions}

The one-loop effective will be $S_\L$ plus the Fradkin-Tseytlin
contribution.
As we mention in the introduction, we are studying the one-loop sigma
model at
tree-level in the world-sheet. That is we
evaluate \ft\ on the sphere, then the world-sheet metric takes the form
$\L dz d\bar z$. Finally the effective action at the one-loop level will
be

\eqn\seff{S_{eff} = S_0 + S_\L
+ {1\over 2\pi}\int d^2z~[\N\Pb^A \N_A\Phi+\Pb^A\Pi^B \N_B \N_A \Phi]\log
\L.}

The beta functions are defined as the $\L$ dependent factor of every
independent coupling of the effective action and, in order to deal with a
theory that is scale independent (such a conformal theory), we need that
this
dependence to be zero. All these couplings are
conformal weights $(1,1)$ constructed out of the products formed in

$$
(\Pi^A, d_\a, J, N^{ab})\times(\Pb^c, \Jb^I),$$
since $\Pb^\a$ is not independent. In fact by varying the action \action\
respect to $d_\a$ we obtain $\Pb^\a=-\Jb^I W_I^\a$. Also we need the
string
equations equations of motion to express $\N\Pb^A$ in terms of the
couplings.
They come form \action\ by varying respect to $\r, \l, \o$ and $Z^M$ as we
will show in the next section.

\newsec{The Vanishing of the Beta Functions}

Now we can write the equations coming from the vanishing of the beta
functions
associated to each independent coupling of the sigma model action. Before
this,
it is convenient to use the set of constraints on the the superfields
given
in \bh. Namely, we use
$H_{\a\b\g}=T_{\a\b}{}^\g=0$ and
$T_{\a\b}{}^a=\g^a_{\a\b}$ besides all the constraints \nilp\ and \hol.
These constraints are fixed using  the scale and one of the local Lorentz
invariances. Note that the remaining local symmetries are those of D=10
N=1
supergravity.
Since the fermionic scale invariance is fixed, the $R_{AB}$
superfield is no longer an invariant curvature, and its connections will
appear
explicitly in the following calculations.
The
first thing we note is that the constraint $T_{a \a}{}^\a=0$ implies that
$\O_a=0$. Also, as it was showed in \bh, the Bianchi identity
$(\N T)_{\a\b\g}{}^a=0$\foot{Recall our notation \bianchi.} determines
$T_{\a ab}=2(\g_{ab})_\a{}^\b \O_\b$. Since it will be used later, it is
useful
to write $R_{AB}$ in terms of other fields
$$R_{ab}=T_{ab}{}^\a \Omega_\a,\quad R_{a \b}=-R_{\b a}=\N_a \Omega_\b,
\quad R_{\a\b}=\N_{(\a}\Omega_{\b)}.$$

Also, it is not difficult to show
that
the Bianchi identity $(\N H)_{ab \a\b}=0$ implies $T_{abc}+H_{abc}=0$.

Now we write the background superfields needed to determine the beta
functions
according to \seff. We use the simplifications derived in the previous
paragraph for the background superfields. After the superspace expansion
described in the section 3, it is not difficult to get

$$
-\eta^{ab}E_{ab}^{(1,1)}=-\half d_{0\a}\Jb^I_0\N^2 W^\a_I
-\half J_0\Jb^I_0\N^2 U_I-{1\over 4}N_0^{ab} \Jb^I_0 \N^2 U_{Iab}$$
$$
-\half
d_{0\a}\Pb^A(\N^a T_{aA}{}^\a+T_{Aa}{}^B T_{Bb}{}^\a\eta^{ab})-\half
\Pi^A\Jb^I_0(\N^a F_{IaA} + T_{Aa}{}^B F_{IB}{}^a)$$
$$
-\half
J_0\Pb^A(\N^a R_{aA}+T_{Aa}{}^B R_B{}^a)-{1\over 4}N_0^{ab}\Pb^A(\N^c
R_{cAab}+T_{Ac}{}^B R_B{}^c{}_{ab})$$
$$
-{1\over 4}\Pi^A\Pb^B(\N^a H_{aBA}+T_{Aa}{}^C
H_{CB}{}^a(-1)^{AB}- T_{Ba}{}^C H_{CA}{}^a+2T_{Aa}{}^c
T_{Bb}{}^d\eta_{cd}\eta^{ab}(-1)^{AB})$$
$$
+{1\over 4}\Pi^{(A}\Pb^{a)}\eta_{ab}(\N^c T_{Ac}{}^b-T_{Ac}{}^C
T_{Cd}{}^b\eta^{cd}+R_{Acd}{}^b\eta^{cd}),$$
where we have to use $\Pb^\a=-\Jb^I W_I^\a$. The remaining terms in \seff\
are obtained from

$$
C^{(0,1)}_{ab}=\half\Jb^I(F_{ab}+W_I^\a T_{\a ab}),$$
$$
D^{(1,0)}_{cd}=\half d_\a T_{cd}{}^\a + \half J R_{cd}
+ {1\over 4} N^{ef} R_{cdef}-\half\Pi^A T_{Acd},$$
$$
D^{(1,0)}_{[\a a]}=J R_{\a a} + \half N^{ef} R_{\a aef},$$
$$
G^{\a(0,1)}_b=\Jb^I\N_b W^\a_I - \Pb^d T_{db}{}^\a,$$
$$
I^{(1,0)}_{J\a}=-d_\b \N_\a W^\b_J - \Pi^A F_{J A\a} + J\N_\a U_J
+\half N^{ef} \N_\a U_{Jef},$$
$$
S^{\a(0.0)}_K=W^\a_K,\quad R^{(0,0)}_{Jab}=\half U_{Jab}.$$

As we said before, we need the superstring equations of motion to
determine
the contribution $\N\Pb^A$ in \seff. After the variation of the \action\
respect to all the world-sheet field, it is not difficult to obtain

$$
\N\Pb_a=\Pi^b\Pb^c T_{abc}+\Pi^\a\Pb^b T_{a\a b}-\Pi^A\Jb^I F_{IA a}
-d_\a\Pb^b T_{ba}{}^\a - d_\a\Jb^I\N_a W^\a_I$$
$$
- J\Pb^b R_{ba}
-\half \Pb^b R_{bacd} + J\Jb^I (\N_a U_I + W^\g_I R_{\g a})
+\half N^{bc}\Jb^I(\N_a U_{Ibc} + W^\g_I R_{\g abc}),$$
and

$$
\N\Pb^\a=-\Jb^I\Pi^A\N_A W^\a_I - d_\b \Jb^I f^{JK}{}_I W^\a_J W^\b_K
+J\Jb^I f^{JK}{}_I W^\a_J U_K +\half N^{ab}\Jb^I f^{JK}{}_I W^\a_J
U_{Kab}.$$

With all this information, we can calculate the terms contributing to
\seff.
Now we write the beta function associated to every independent coupling
of the effective action. They are separated in two groups. The first group
of
equations is the SUGRA sector

\eqn\sugra{\eqalign{
&\N_{a} \N_{b} \Phi -\half {\cal R}_{ab} +
\half T_{ca}{}^\a T_{b\a}{}^c = 0,\cr
&\N^c T_{abc} - 2 T_{abc} \N^c \Phi + 2 T_{ab}{}^\a \N_\a \Phi = 0,\cr
&\N^b T_{\a ba} + R_{\a cda} \eta^{cd} + T_{ab}{}^\b \g^b_{\b\a}
+ 4 \N_a \N_\a \Phi = 0,\cr
&\N^b T_{ba}{}^\a - T_{bc}{}^\a T_a{}^{bc} + 2 T_{ab}{}^\a \N^b \Phi =
0,\cr
&\N^b R_{ba} - T_{abc} R^{bc} - T_{ab}{}^\a R_\a{}^b
+ 2 R_{ab} \N^b \Phi = 0,\cr
&\N^b R_{bafg} - T_{abc} R^{bc}{}_{fg} - T_{ab}{}^\a R_\a{}^b{}_{fg}
+ 2 R_{abfg} \N^b \Phi = 0,\cr}}
where ${\cal R}_{ab}=R^c{}_{acb}$ is the Ricci tensor which is not 
symmetric. The second
group of equations is the SYM sector

\eqn\sym{\eqalign{
&\g^a_{\a\b} \N_a W^\b_I = 2(\g^a_{\a\b} \N_a \Phi) W^\b_I
+ 2 \N_\a (\N_\b\Phi W^\b_I),\cr
&\N^b F_{Iba} - 2 f^{JK}{}_I F_{J a\a} W^\a_K + 2 F_{Iab} \N^b \Phi - 2
\N_a (\N_\a\Phi W^\a_I)\cr
&+ \half W^\a_I ( \N^b T_{\a ba} + T_{ab}{}^\b \g^b_{\b\a} + R_{\a cda}
\eta^{cd} ) = 0,\cr
&\N^2 W^\a_I - F_I^{ab} T_{ab}{}^\a + 2 (\N_a W^\a_I)\N^a \Phi -
2 f^{JK}{}_I ( \N_\b W^\a_J +  W^\a_J \N_\b \Phi ) W^\b_K=0,\cr
&\N^2 U_I -  F_I^{ab} R_{ab} +  W^\b_I ( 2 R_{a\b} \N^a \Phi - \N^a
R_{a\b} )
+ 2 R_{a\a} \N^a W^\a_I\cr
&- 2 (\N_a U_I) \N^a \Phi + 2 f^{JK}{}_I ( \N_\a U_J - U_J \N_\a \Phi )
W^\a_K = 0,\cr
&\N^2 U_{Ifg} -  F_I^{ab} R_{abfg} + W^\b_I ( 2 R_{a\b fg} \N^a \Phi -
\N^a R_{a\b fg} )
+ 2 R_{a\a fg} \N^a W^\a_I\cr
&- 2 (\N_a U_{Ifg}) \N^a \Phi + 2 f^{JK}{}_I
( \N_\a U_{Jfg} - U_{Jfg} \N_\a \Phi )
W^\a_K + f^{JK}{}_I U_{Ja[f} U_{Kg]}{}^a = 0.\cr}}

We need to verify that the constraints given in \bh\ plus the use of the
Bianchi identities \bianchi\ allow to verify the equations \sugra\ and
\sym. It
is a tedious but direct job, we follow the idea of the calculation done in
the
reference \ref\adr{
J.~J.~Atick, A.~Dhar and B.~Ratra,
``Superspace Formulation of Ten-dimensional N=1 Supergravity Coupled to
N=1
Super Yang-Mills Theory,''
Phys.\ Rev.\ D {\bf 33}, 2824 (1986).}.

\subsec{The SUGRA sector}

Now it will be shown that the SUGRA set of equations \sugra\ are implied
by
the use of the constraints \nilp, \hol\ and the use of the Bianchi
identities.
First we note that it is satisfied

\eqn\ttrr{\g^b_{\a\b} T_{ba}{}^\b = 8 R_{a\a}.}
To show this, we see that $R_{\a abc}$ can be written, by using the
Bianchi identity $(\N T)_{\a ab}{}^c=0$, as

$$
R_{\a abc} = T_{a[b}{}^\b (\g_{c]})_{\b\a} - 2 (\g_{bc})_\a{}^\b
R_{a\b},$$
we plug this into the Bianchi identity $(\N T)_{a\a\b}{}^\b=0$ to verify
\ttrr.
 Now it is trivial to show that the third equation in \sugra\ is
satisfied.

Now we will show that the second equation in \sugra\ is also satisfied.
Consider the Bianchi identity $(\N T)_{abc}{}^c=0$, it implies

$$
\N^c T_{abc} + 4 \O_\a T_{ab}{}^\a-16\O_\a\g^{\a\b}_{[a} R_{b]\b}
+\eta^{cd}(R_{acbd}-R_{bdac})=0.$$
From the Bianchi identity $(\N T)_{\a ab}{}^\b=0$ one can obtain

\eqn\rabcd{
R_{abcd}=-{1\over 8}\N_\a ( (\g_{cd})_\b{}^\a T_{ab}{}^\b )
+ {1\over 8} (\g_{cd})_\b{}^\a T_{\a[a}{}^e T_{b]e}{}^\b,}
which allows to write $\eta^{cd}(R_{acbd}-R_{bdac})$ and, after plugging
it
into the above equation one obtains

$$
\N^c T_{abc} + \g^{\a\b}_a \N_\a R_{b\b} -  \g^{\a\b}_b \N_\a R_{a\b}
+ 2 \O_\a (\g_a\g_{bc}-\g_b\g_{ac})^{\b\a} R^c{}_\b=0,$$
finally if one uses $R_{a\a}=\N_a \O_\a$ and commutes the derivatives in
$\N_\a R_{b\b}$ and in $\N_\a R_{a\b}$ one can arrive to the second
equation
in \sugra. We can verify the first equation in \sugra\ by constructing the
Ricci tensor ${\cal R}_{ab}=R^c{}_{acb}$ from \rabcd.

It remains to verify the last three
equations in \sugra. Note that the fourth and the fifth equations are
equivalent by recalling the relation $R_{ab}=T_{ab}{}^\a \Omega_\a$.
Without using the previous equations, the derivation of the fourth
equation is
more involved. Using the Bianchi indentities $(\N R)_{[\a ab]\b}{}^\g=0$,
$(\N T)_{c\a\b}{}^\g=0$ and
$(\gamma_a)^{\a\b}R_{\a\b\g}{}^\d=-2(\gamma_a)^{\a\b}R_{\g\a\b}{}^\d$,
that follows from $(\N T)_{\a\b\g}{}^\d=0$ we show that

$$(\gamma^a)^{\a\b}(\N_\a R_{ab\b}{}^\g - T_{\a[a}{}^e R_{b]e\b}{}^\g) -
8T_{b}{}^{cd}T_{cd}{}^\g + 2T_{eb}{}^\g \N^e\Phi + 8\N^e T_{eb}{}^\g$$
$$
-{1\over 8}(\g^a)^{\a\b}(\g^{cd})_\r{}^\g R_{\a\b cd} T_{ab}{}^\r=0,$$
and after working out the first and the last terms by using \rabcd\ and
the Bianchi identity $(\N T)_{\a\b a}{}^b=0$, we see
that they give exactly the
remaining terms, proving the fourth equation.
Finally the last equation is satisfied by using \rabcd\ and
the remaining equations in \sugra.

We can see that the first equation is the graviton equation of motion, the
second is the equation for the antisymmetric tensor.
Contracting the third equation with $(\gamma^a)^{\a\b}$ we get the
dilatino equation of motion and contracting with $(\gamma^{ac})_\b{}^\a$ we 
get the gravitino equation.\foot{Remember that at linearized level 
$T_{ab}{}^\a \approx \p_{[a}\psi_{b]}{}^\a$, where $\psi_a{}^\a$ is the 
gravitino.} 
The last three equations are redundant, they can be obtained from the others.
These results prove the claim in \bh\ that the classical BRST invariance is
equivalent to quantum conformal invariance at 1-loop level.

\subsec{The SYM sector}

The verification of the SYM equations is the following. Let us first
consider
the first equation in \sym. To verify that it is implied by the Bianchi
identities it is necessary to relate the field strength $F$ with the
superfield $U$. The Bianchi identity $(\N F)_{a\a\b}=0$ implies

$$
U_I+\O_\a W^\a_I=0,$$
$$
F_{Iab}=U_{Iab}+2(\g_{ab})_\a{}^\b\O_\b W^\a_I.$$
Besides, it will be necessary to know the spinorial derivative on this
superfield. The Bianchi identity $(\N F)_{\a ab}=0$ determines

$$
\N_\a F_{Iab} = -\N_{[a} F_{Ib]\a} + T_{\a[a}{}^c F_{Ib]c} -
T_{ab}{}^c F_{Ic\a}.$$
The gluino equation can be obtained from the identity

$$
\{\N_\a,\N_\b\}W^\b_I=-T_{\a\b}{}^a \N_a W^\b_I + R_{\a\g\b}{}^\g
W^\b_I,$$
and using the constraint equation for $\N_\a W^\b_I$ in \hol. It is also
necessary to determine $R_{\a\g ab}(\g^{ab})_\b{}^\g$ from the Bianchi
identity $(\N T)_{\a\b a}{}^b=0$:

$$
R_{\a\g
ab}(\g^{ab})_\b{}^\g=-180\N_\a\O_\b+2(\g^{ab})_\a{}^\g(\g_{ab})_\b{}^\d
\N_\d\O_\g-\g^{abc}_{\a\b} T_{abc}-384\O_\a\O_\b,$$
where we have used the identity $(\g^{ab})_\a{}^\r(\g_{ab})_\b{}^\s \O_\r
\O_\s=6\O_\a\O_\b$.
Doing all this we obtain

$$
-{7\over 2} \g^a_{\a\b}\N_a W^\b_I -28\N_\a U_I -[R_{\a\b}+\half
(\g^{ab})_\a{}^\r(\g_{ab})_\b{}^\s R_{\r\s}]W^\b_I=0,$$
where $R_{\a\b}=\N_{(\a} \O_{\b)}$. The Bianchi identity
$R_{(\a\b\g)}{}^\g=0$ allows us to finally write

$$
\g^a_{\a\b}\N_a W^\b_I=-8(\N_\a U_I+ R_{\a\b} W^\b_I).$$
which takes the form of the first equation in \sym\ if we recall \dpom.

Now we satisfy the second equation in \sym. The idea is to start with the
knowledge of the commutator $[\N_b,\N_\a] W^\b_I$, which can be obtained
from
\nn, and multiply by $(\g^{ab})_\b{}^\a$, and use the first equation in
\sym together with the third equation in \sugra. After a tedious,
but direct calculation, the second equation can be
showed to be satisfied. Similarly for the remaining equations in the
SYM sector can be verified as consequence of the first two equations in
\sym. The third equation is obtained by applying $(\g^b)^{\a\g}\N_b$ in
the
first equation in \sym, while the last two equations can be obtained by
acting with $\N_\b$ on the third
equation and after commuting it with $\N^2$. Then, the fourth
equation is obtained by contracting with $\d^\b_\a$ and we get
the fifth equation and by multiplying with $(\g_{ef})_\a{}^\b$.

\vskip 15pt
{\bf Acknowledgements:} We would like to thank Nathan Berkovits, Jim
Gates, Paul Howe, Paolo Pasti, Dimitri Sorokin, Mario Tonin for 
useful comments and suggestions. OC would like to thank the INFN 
for a post-doctoral fellowship and FONDECYT grant
3000026 for partial financial support. The work of BCV is supported by
FAPESP grant 00/02230-3.

\listrefs

\end